\documentclass[12pt]{iopart}
\usepackage{iopams}
\usepackage{graphicx}
\usepackage{epsfig} 
 
\begin{document}

\title{Efficient optical quantum information processing}

\author{W.J.\ Munro$^\dag$\footnote[3]{bill.munro@hp.com}, Kae
  Nemoto$^\ddag$, T.P.\ Spiller$^\dag$, S.D.\ Barrett$^\dag$,
  Pieter Kok$^\dag$, and R.G.\ Beausoleil$^\ast$} 

\address{\dag\ Quantum Information Processing Group, Hewlett-Packard
  Laboratories, Filton Road, Stoke Gifford, Bristol BS34 8QZ, United
  Kingdom} 
\address{\ddag\ Quantum Information Science Group, National Institute
  of Informatics, 2-1-2 Hitotsubashi, Chiyoda-ku, Tokyo 101-8430,
  Japan} 
\address{$\ast$\ Hewlett-Packard Laboratories, 13837 175$^\textrm{th}$
  Pl.\ NE, Redmond, WA 98052--2180, USA} 

\begin{abstract}
 Quantum information offers the promise of being able to perform
 certain communication and computation tasks that cannot be done with
 conventional information technology (IT). Optical Quantum Information Processing (QIP) holds
 particular appeal, since it offers the prospect of communicating and
 computing with the same type of qubit. Linear optical techniques have
 been shown to be scalable, but the corresponding quantum computing
 circuits need many auxiliary resources. Here we present an
 alternative approach to optical QIP, based on the use of weak
 cross-Kerr nonlinearities and homodyne measurements. We show how this
 approach provides the fundamental building blocks for highly
 efficient non-absorbing single photon number resolving detectors, two
 qubit parity detectors, Bell state measurements and finally near
 deterministic control-not (CNOT) gates. These are essential QIP devices.
\end{abstract}




\section{Introduction}

It has been known for a number of years that processing information
quantum mechanically enables certain communication and computation
tasks that cannot be performed with conventional Information
Technology (IT). The list of applications continues to expand, and
there are extensive experimental efforts in many fields to realise the
necessary building blocks for Quantum Information Processing (QIP)
devices. One very appealing route (certainly in the short term when a
pragmatic focus is on few-qubit applications) is that of optical
QIP. In particular for quantum communication the qubits of choice are 
optical systems, since they can span long distances with minimal
decoherence. In order to circumvent interconversion of the qubit
species we need to process the optical qubits using optical circuits.

Optical QIP is currently a very active research area, both
theoretically and experimentally. The work of Knill, Laflamme and
Milburn (KLM) has shown that in principle universal quantum
computation is possible with linear optics \cite{KLM}, and there have
been a number of recent experimental demonstrations of its gate
components \cite{Pittman03,OBrien03,Gasparoni04}. However, due to the   
probabilistic nature of gates in linear optical QIP, it is practically 
rather inefficient (in terms of photon resources) to implement
\cite{scheel03,scheel04,Eisert04}. 

Strong Kerr non-linearities are able to effectively mediate an
interaction directly between photonic qubits\cite{Vitali00,Ottaviani03}. 
This would realise
deterministic quantum gates and thus efficient optical QIP. In
practice, however, such non-linearities are not available. On the
other hand, much smaller non-linearities can be generated, for
example, with electromagnetically induced transparencies 
(EIT)\cite{Imamoglu96,Paternotro02,munro03}. In
this paper, we show that with modest additional optical resources 
these small non-linearities provide the building blocks for efficient
optical  QIP \cite{grang98,nemoto04,barrett04,paris00,DAriano00}. We
present a highly efficient non-absorbing single photon number resolving
detector, a nondestructive two qubit parity detector, a nondestructive
Bell state measurement, and a near deterministic controlled-not (CNOT) gate.

\section{Quantum non-demolition detectors}

\begin{figure}[!htb]
\center{\includegraphics[scale=0.6]{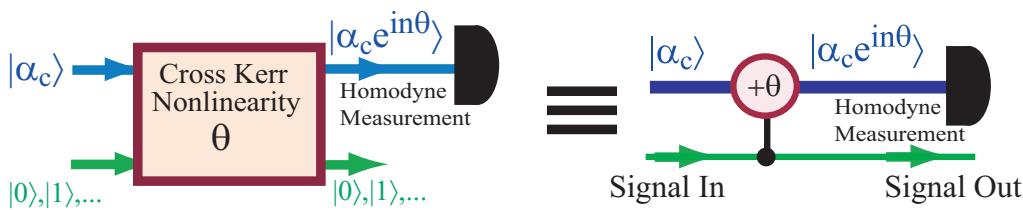}}
\caption{Schematic diagram of a photon resolving detector based on a
  cross-Kerr Non-linearity and a homodyne measurement. The two inputs
  are a Fock state $|n_a\rangle$ (with $n_a=0,1,$..) in the signal
  mode $a$ and a coherent state with real amplitude $\alpha_c$ in the
  probe mode $c$. The presence of photons in mode $a$ causes a phase
  shift on the coherent state $|\alpha_c\rangle$ directly proportional
  to $n_a$ which can be determined with a momentum quadrature
  measurement.}\label{detectorhomodyne}
\end{figure}

Before we discuss the construction of efficient quantum gates using
weak non-linearities, let us first review the construction of a photon
number quantum non-demolition (QND) measurement using a cross-Kerr non-linearity \cite{Milburn
  1984,imot85,munro03}. The cross-Kerr non-linearity has a Hamiltonian
of the form: 
\begin{eqnarray}
H_{QND}= \hbar \chi a^\dagger a c^\dagger c
\end{eqnarray}
where the signal (probe) mode has the creation and annihilation
operators given by $a^\dagger, a$ ($c^\dagger, c$) respectively
and $\chi$ is the strength of the non-linearity. If the signal field
contains $n_a$ photons and the probe field is in an initial coherent
state with amplitude $\alpha_c$, the cross-Kerr non-linearity
causes the combined system to evolve as
\begin{eqnarray}\label{phase-shift}
|\Psi(t)\rangle_{out}
&=&e^{i \chi t a^\dagger a c^\dagger c} |n_a\rangle|\alpha_c\rangle
=|n_a\rangle|\alpha_c e^{i n_a \theta}\rangle .
\end{eqnarray}
where $\theta=\chi t$ with $t$ being the interaction time for the signal 
and probe modes with the non-linear material. The Fock state
$|n_a\rangle$ is unaffected by the interaction with the cross-Kerr
non-linearity but the coherent state $|\alpha_c\rangle$ picks up a
phase shift directly proportional to the number of photons $n_a$ in
the signal $|n_a\rangle$ state. If we could measure this phase shift
we could then infer the number of photons in the signal mode $a$. This
can be achieved simply with a homodyne measurement (depicted
schematically in figure (\ref{detectorhomodyne})).
The homodyne apparatus allows measurement of the quadrature operator
$x(\phi)=c e^{i \phi} +c^\dagger e^{-i \phi}$ with expectation value
\begin{eqnarray}
\langle x(\phi)\rangle &=& 2 {\rm Re}\left[\alpha_c\right] \cos
\delta +i 2 {\rm Im}\left[\alpha_c \right]\sin \delta
\end{eqnarray}
where $\delta=\phi+ n_a \theta$ and $\phi$ the phase of the 
local oscillator. For a real initial $\alpha_c$, 
a highly efficient homodyne measurement of the position $X=a+a^\dagger$ 
or momentum $iY=a-a^\dagger$ quadratures yield the expectation values
\begin{eqnarray}
\langle X \rangle &=& 2 \alpha_c \cos \left(n_a \theta \right)
\;\;\;\;\;\;\;\;\;\;\;\; 
\langle Y \rangle = 2 \alpha_c \sin \left(n_a \theta \right)
\end{eqnarray}
with  a unit variance. For the momentum quadrature this gives a
signal-to-noise ratio ${\rm SNR}_Y=2 \alpha_c \sin \left(n_a \theta
\right)$ which should be much greater than unity for the different
$n_a$ inputs to be distinguished. In more detail, if the inputs in
mode $a$ are the Fock state $|0\rangle$ or $|1\rangle$, the respective
outputs of the probe mode $c$ are the coherent states
$|\alpha_c\rangle$ or $|\alpha_c e^{i \theta}\rangle$. The probability
of misidentifying these states is given by 
\begin{eqnarray}
P_{\rm error}= \frac{1}{2} {\rm Erfc} \left[ \alpha_c  \sin \theta /
  \sqrt{2}   \right] =\frac{1}{2} {\rm Erfc} \left[ {\rm SNR}_Y / 2
  \sqrt{2}   \right] .
\end{eqnarray}
A signal to noise ratio of ${\rm SNR}_Y=6$ would thus give $P_{\rm
  error} \sim 10^{-3}$. To achieve the necessary phase shift we
require $\alpha_c \sin \theta \approx 3$ which can be achieved with a
small non-linearity $\theta$ as long as the probe beam is intense
enough. These results so far indicate that with a weak cross-Kerr
non-linearity it is possible to build a high efficiency photon number
resolving detector that does not absorb the photon from the signal
mode. In reference \cite{munro03} we discuss an example of how this
level of non-linearity could be achieved through electromagnetically
induced transparency (EIT), and give some details for this approach
with NV-diamond systems.  

\begin{figure}[!htb]
\center{\includegraphics[scale=0.7]{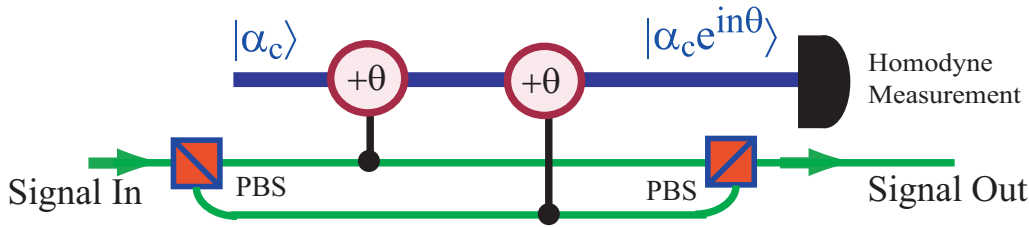}}
\caption{Schematic diagram of a polarization-preserving photon
number quantum non-demolition detector based on a pair of identical
cross-Kerr optical non-linearities. The signal mode is a Fock state
with an unknown polarization is converted into which path qubits 
by a polarizing beam splitter (square box). The phase shift
applied to the probe mode is proportional to $n_a$, independent of
the polarization of the signal mode.}
\label{qnd_pp}
\end{figure}

In many optical quantum computation tasks our information is not
encoded in photon number but polarization instead. When our
information  encoding is polarization based there are two separate
detection tasks  that we need to be able to perform. The first and
simplest is just to  determine for instance whether the polarization
is in one of the basis  states $|H\rangle$ or $|V\rangle$. This can be
achieved by converting the  polarization information to ``which path''
information on a polarizing  beam-splitter. The ``which path''
information is photon number encoded in each  path and hence a QND
photon number measurement of each path will determine  which
polarization basis state the photon was originally in. The second task
(one that is critically important for error correction codes in
optics) is to determine whether our single-photon polarization-encoded
qubit is present or not. That is, for the optical field under
consideration we want to determine whether it contains a photon or
not. If it does contain a photon, we do not want to destroy the
information in its polarization state. This can be
achieved by first converting the  polarization qubit to a ``which
path'' qubit. Each path then interacts with a weak cross-Kerr
non-linearity $\theta$ with the same shared probe beam
(Figure~\ref{qnd_pp}). If the photon is present in either path of this
signal beam it induces a phase shift $\theta$ on the probe beam;
however, with  this configuration it is not possible to determine
which path induced the phase shift. This allows the preservation of
the ``which path''  and hence polarization information,

\section{A Two Qubit Parity Gate}

Now that we have discussed the basic operation of a single photon
quantum  non-demolition detector, it is worthwhile asking whether this
detection  concept can be applied to several qubits. Basically  if we
want to  perform a more ``generalized'' type of measurement between
different photonic  qubits, we could delay the homodyne measurement,
instead having  the probe beam interact with several cross-Kerr
non-linearities where the  signal mode is different in each case. The
different signal modes could be from  separate photonic qubits. The
probe beam measurement then occurs after  all these interactions in a
collective way which could for instance allow  a nondestructive
detection that distinguishes superpositions and mixtures  of the
states $|HH\rangle$ and $|VV\rangle$ from $|HV\rangle$ and
$|VH\rangle$.  The key here is that we could have no net phase shift
on the $|HH\rangle$  and $|VV\rangle$ terms while having a phase shift
on the $|HV\rangle$ and $|VH\rangle$ terms. We will call this
generalization a {\it two qubit  polarization parity QND gate}.

\begin{figure}[!htb]
\center{\includegraphics[scale=0.7]{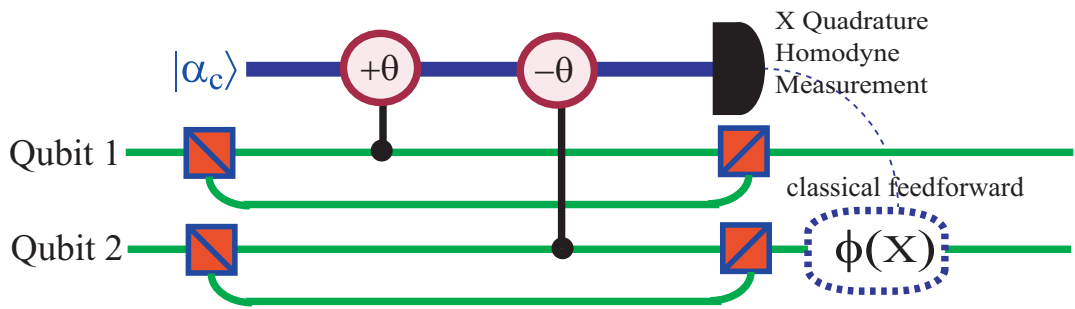}} 
\caption{Schematic diagram of a two qubit polarization QND detector
  that distinguishes superpositions and mixtures of the states
  $|HH\rangle$ and $|VV\rangle$ from $|HV\rangle$ and $|VH\rangle$
  using several cross-Kerr non-linearities and a coherent laser probe
  beam $|\alpha\rangle$. The scheme works by first splitting each
  polarization qubit into a which path qubit on a polarizing
  beam-splitter. The action of the first cross-Kerr non-linearity puts
  a phase shift $\theta$ on the probe beam only if a photon was
  present in that mode. The second cross-Kerr non-linearity put a
  phase shift $-\theta$ on the probe beam only if a photon was present
  in that mode. After the non-linear interactions the which path qubit
  are converted back to polarization encoded qubits. The probe beam
  only picks up a phase shift if the states $|HV\rangle$ and/or
  $|VH\rangle$ were present and hence the appropriate homodyne
  measurement allows the states $|HH\rangle$ and $|VV\rangle$ to be
  distinguished from $|HV\rangle$ and $|VH\rangle$. The two qubit
  polarization QND gate thus acts like a parity checking device. If we
  consider that the input state of the two polarization qubit is
  $|HH\rangle+|HV\rangle+|VH\rangle+|VV\rangle$ then after the parity
  gate we have conditioned on an $X$ homodyne measurement either to
  the state $|HH\rangle+|VV\rangle$  or  to $e^{i \phi (X)}
  |HV\rangle+e^{-i \phi (X)}|VH\rangle$ where $\phi(X)$ is a phase
  shift dependent on the exact result of the homodyne measurement. A
  simple phase shift achieved via classical feed-forward then allows
  this second state to be transformed to the first if we wish.} 
\label{fig-qnd-parity}
\end{figure}

Let us now discuss the operation of this parity QND gate. Consider 
a general two qubit state which can be written as 
$|\Psi_{12}\rangle= \beta_0 |H H \rangle +  \beta_1 |H V \rangle 
+ \beta_2 |V H \rangle + \beta_3 |V V \rangle$. This may be separable 
or entangled depending on the choices of $\beta_i$. As shown
in Figure (\ref{fig-qnd-parity}), these qubits are individually split on 
polarizing beam-splitters (PBS) into spatial encoded qubits which then 
interact with separate weak cross-Kerr non-linearities.  The action 
of the PBS's and cross-Kerr non-linearities evolves the combined system 
of photonic qubits and probe beam to  
\begin{eqnarray}
|\psi\rangle_{T}&=&\left[\beta_0 |H H \rangle +\beta_3|V V \rangle
  \right] |\alpha_c\rangle_p + \beta_1 |H  V \rangle |\alpha_c e^{i
  \theta}\rangle_p +\beta_2 |V H \rangle |\alpha_c e^{- i
  \theta}\rangle_p \nonumber 
\end{eqnarray}
It is now obvious that the $|H H \rangle$ and $|V V \rangle$ terms pick 
up no phase shift and remain coherent with respect to each other while 
the $|H  V \rangle$ and $|V  H \rangle$ pick up opposite sign phase 
shift $\theta$ which could allow them to be distinguished by a general 
homodyne/heterodyne measurement. We thus need to perform a measurement
that does not allow the sign of the phase shift to be determined. With
$\alpha_c$ real an $X$ homodyne measurement achieves this by
projecting the probe beam to the position quadrature eigenstate
$|X\rangle \langle X| $\cite{barrett04}. The resulting two photonic qubit state is then 
\begin{eqnarray}\label{cats}
 &&|\psi_X\rangle_{T}={\it f}(X,\alpha_c)\left[\beta_0 |H H \rangle
    +\beta_3|V V \rangle \right]  \\ 
 &&\;\;\;\;\;\;\;\;\;\;+ {\it f}(X,\alpha_c \cos \theta) \left[ \beta_1
    e^{i \phi(X)} |H  V \rangle+ \beta_2 e^{-i \phi(X)}|V  H \rangle
    \right] \nonumber 
\end{eqnarray}
where 
\begin{eqnarray}
 {\it f}(x,\beta)&=&\exp \left[-\frac{1}{4}
   \left(x-2\beta\right)^2\right]/(2 \pi)^{1/4}\\ 
 \phi(X)&=& \alpha_c \sin \theta ( x  - 2 \alpha_c  \cos \theta) {\rm mod}
   2\pi\, . 
\end{eqnarray}
We observe that ${\it f}(X,\alpha)$ and ${\it f}(X,\alpha \cos\theta)$ are 
two Gaussian  curves with the mid point between the peaks located at 
$X_0=\alpha_c \left[1+\cos \theta \right]$ and the peaks separated by 
a distance $X_d=2 \alpha_c \left[1-\cos  \theta \right]$. As long as 
this difference is large $X_d\sim \alpha_c \theta^2 \gg 1$, then there is 
little overlap between these curves. For an $X$ homodyne $X>X_0$ our solution 
(\ref{cats}) collapses to 
\begin{eqnarray}\label{even-parity}
|\psi_{X>X_0}\rangle_{T}\sim \beta_0 |H H \rangle +\beta_3 |V  V \rangle
\end{eqnarray}
while for $X<X_0$ we have
\begin{eqnarray}\label{odd-parity}
|\psi_{X<X_0}\rangle_{T}\sim \beta_1 e^{i \phi(X)} |H  V \rangle+\beta_2 e^{-i \phi(X)}|V H \rangle 
\end{eqnarray}
The action of this two mode polarization non-demolition parity gate is
clear: It splits the even parity terms (\ref{even-parity}) nearly  
deterministically from the odd parity cases (\ref{odd-parity}). 

Above we have chosen to call the even parity state \{$|HH\rangle,
|VV\rangle$\} and the odd parity states \{$|HV\rangle, |VH\rangle$\},
but this is an arbitrary choice primarily dependent on the form/type
of PBS used to convert the polarization encoded qubits to which path
encoded qubits. Any other choice is also acceptable and it does not
have to be symmetric between the two qubits. 

Our solution in Eqn (\ref{odd-parity}) depends on the value of the
measured quadrature $X$. Simple local rotations using phase shifters
dependent on the measurement result $X$ can be performed via a feed
forward process to transform this state to $\beta_1 |H \rangle_a |V
\rangle_b + \beta_2 |V \rangle_a |H \rangle_b$ which is independent of
$X$. This does mean our homodyne measurement must be accurate enough
such that we can determine $\phi(X)$ precisely, otherwise this
unwanted phase factor cannot be undone. By this we mean that the
uncertainty in the $X$ quadrature homodyne measurement must be much
less than $2 \pi / \alpha_c \sin \theta$ and this can generally be
achieved by ensuring that the strength of the local oscillator is much
more intense than the probe mode.  

In the above solutions (\ref{even-parity}) and (\ref{odd-parity}) we
have used the the approximate symbol $\sim$ as there is a small but
finite probability that the state (\ref{even-parity}) can occur for
$X<X_0$ and vice versa. The probability of this error occurring is
given by 
\begin{eqnarray}
 P_{\rm error}=\frac{1}{2} {\rm Erfc} [X_d/{2\sqrt 2}]
\end{eqnarray}
which is less than $10^{-4}$ when the distance $X_d \sim \alpha_c
\theta^2 > 8$. This shows that it is possible to operate in the regime
of weak cross-Kerr non-linearities ($\theta \ll \pi$) and achieve an
effectively deterministic parity measurement.

\section{Bell state measurements}

These effectively deterministic nondestructive parity measurements are
critically important in optical quantum information processing as they
naturally allow an efficient and deterministic Bell state measurement
to be implemented. Bell state measurements are known to be one of the
tools and mechanism required in quantum computation and
communication. The four Bell states can be written as 
\begin{eqnarray}
 |\Psi^{\pm}\rangle &\equiv& \frac{1}{\sqrt{2}} \left( |H,V\rangle \pm
 |V,H\rangle \right) \qquad |\Phi^{\pm}\rangle \equiv
 |\frac{1}{\sqrt{2}} \left( |H,H\rangle \pm |V,V\rangle \right) 
\end{eqnarray}
and we can now see why the parity gate can form the basis of a Bell
state detector. The parity gate distinguishes states within the even
parity $|H,H\rangle$ and $|V,V\rangle$ subspace from the odd parity
$|H,V\rangle$ and $|V,H\rangle$ subspace. Hence one application of the
parity detector distinguishes two of the Bell states
$|\Phi^{\pm}\rangle$ from the $|\Psi^{\pm}\rangle$ ones without
destroying them. Similarly if we replace the polarizing beam-splitter
in the parity gates with 45 degree PBS then the parity gate will allow
us to distinguish the $|\Phi^+\rangle,|\Psi^+\rangle$ Bell states from
$|\Phi^-\rangle,|\Psi^-\rangle$ ones.  

\begin{figure}[!htb]
\center{\includegraphics[scale=0.6]{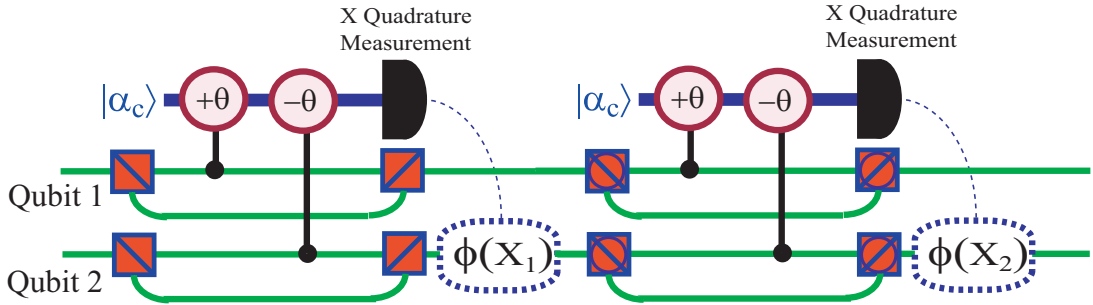} }
\caption{Schematic diagram of a non-destructive Bell state measurement
  composed of two QND parity detectors. The first parity gates uses
  the standard PBS and distinguishes the $|\Phi^{\pm}\rangle$ Bell
  states from the $|\Psi^{\pm}\rangle$ ones. An even parity results
  for this first parity gate indicates the present of the
  $|\Phi^{\pm}\rangle$ Bell states while an odd parity result
  indicates the present of the $|\Psi^{\pm}\rangle$ states. For this
  odd parity result a local operation on the second qubit is required
  to remove the $\phi (X_1)$ phase shifted induced from the
  measurement. Once this correction is done the second parity gate can
  be applied. This gate is similar to the first one but has 45 deg
  PBS's (square box with circle inside) instead of the normal
  PBS's. The 45 deg PBS's operate in the  \{H+V,H-V\} basis. An even
  parity result indicates the presence of the
  $|\Phi^+\rangle,|\Psi^+\rangle$ Bell states while an odd parity
  result indicates the presence of the $|\Phi^-\rangle,|\Psi^-\rangle$
  Bell states. Again a phase correction $\phi (X_2)$ in the
  \{H+V,H-V\} basis is needed for the odd parity result to remove the
  unwanted phase shift. } 
\label{fig-qnd-bell}
\end{figure}

Since both of these detectors are nondestructive on the qubits and
select different pairs of Bell states, they allow the natural
construction of a Bell state detector (depicted in Fig
(\ref{fig-qnd-bell})). From each parity measurements we get one of
information indicating whether the parity was even or odd and so from
both parity measurements we end up with four possible results (even,
even), (even, odd), (odd, even) and (odd, odd). This is enough to
uniquely identify all the Bell states as the $|\Phi^+\rangle$ gives
the result (even, even), $ |\Phi^-\rangle$  (even, odd),  $
|\Psi^+\rangle$  (odd, even) and $ |\Psi^-\rangle$  (odd, odd). It is
important that after an odd parity measurement result that we remove
the unwanted phase factors that have arisen. This need to be done in
the same basis as the PBS in the particular parity gate. For instance
for an odd parity results giving $X=X_1$ on the first parity gate a
phase shift $\phi (X_1)$ needs to be removed in the PBS $\{H,V\}$
basis. Similarly for a odd parity results giving $X=X_2$ on the second
parity gate a phase shift $\phi (X_2)$ needs to be removed in the PBS
$\{H+V,H-V\}$ basis 

So far we have shown how it is possible using linear elements, weak
cross-Kerr non-linearities and homodyne measurements to create a wide
range of high efficiency quantum detectors and gates that can perform
task ranging from photon number discrimination to Bell state
measurements. This is all achieved non-destructively on the photonic
qubits and hence provides a critical set of tools extremely useful for
single photon quantum computation and communication. With these tools
universal quantum computation can be achieved using the ideas and
techniques originally proposed by KLM.   

\section{A resource efficient CNOT gate}

\begin{figure}[!htb]
\center{\includegraphics[scale=0.5]{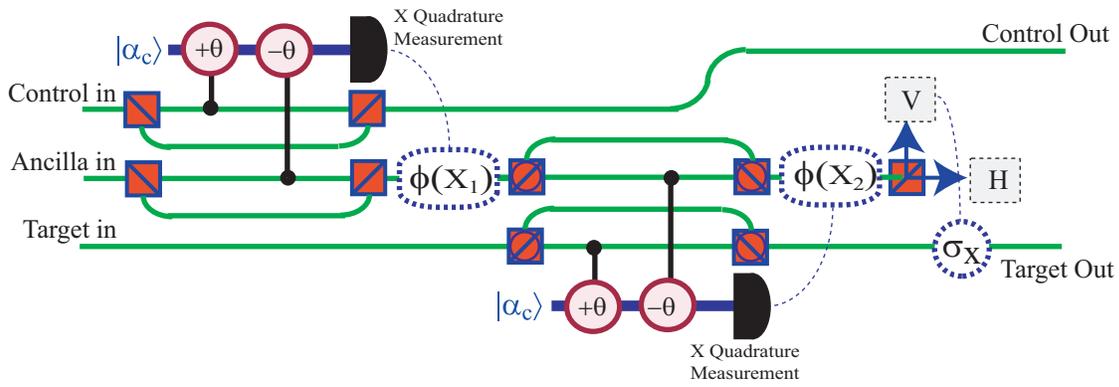}}
\caption{Schematic diagram of a near deterministic CNOT composed of two 
parity gates (one with PBS in the \{H,V\} basis and one with 
PBS in the \{H+V,H-V\} basis), one ancilla qubit prepared initially 
as $|H\rangle+|V\rangle$, a polarisation determining photon number 
QND measurement and classical feed-forward elements.}
\label{fig-qnd-cnot}
\end{figure}

The parity gate and Bell state detector has shown how versatile the weak 
non-linearities and homodyne conditioning measurements are. Both of these 
gates/detectors can be used to induce two qubit operations and hence are 
all they are necessary with single qubit operation and single photon 
measurements to perform universal quantum computation. The parity and Bell 
state gates are not the typical two qubits gates that one generally 
considers in the standard quantum computational models. The typical two 
qubit gate generally considered is the CNOT gate. This can be constructed 
from two parity gates (like the Bell state detector) but it also requires 
an ancilla qubit. This CNOT gate is depicted schematically shown in 
Fig (\ref{fig-qnd-cnot})) and operates as follows.

Assume that our control and target qubits are initially prepared as
$c_0 |H \rangle_c + c_1 |V \rangle_c $ and $d_0 |H \rangle_t + d_1 |V
\rangle_t$. With an ancilla qubit prepared as $|H\rangle_a +|V
\rangle_a$ the action of the first parity gate on the control and
ancilla qubits (with appropriate phase corrections for the odd parity
result) conditions the system to
\begin{eqnarray}
 \left[c_0 |HH \rangle_{ca} + c_1 |VV \rangle_{ca}\right]\otimes
 \left[d_0 |H \rangle_t + d_1 |V \rangle_t\right] 
\end{eqnarray}
The action of the second parity gate (using 45 deg PBS's instead of
normal PBS's) on the ancilla qubit and target qubit conditions the
three qubit system to  
\begin{eqnarray}
 \left\{c_0 |H \rangle_c - c_1 |V \rangle_c\right\} (d_0-d_1)|\bar
 D,\bar D\rangle_{at}+ \left\{c_0 |H \rangle_c + c_1 |V
 \rangle_c\right\} (d_0+d_1)|D,D\rangle_{at} 
\end{eqnarray}
where $|D\rangle=|H\rangle+|V\rangle$, $|\bar
D\rangle=|H\rangle-|V\rangle$ and for the odd parity measurement
result $X<X_0$ the usual phase correction is applied. Also for this
odd parity result a bit flip is applied to the ancilla qubit and a
sign flip $|V \rangle_c\rightarrow -|V \rangle_c$ on the control
qubit. Once this operation have been performed the ancilla mode is
measured in the $\{H,V\}$ basis using QND photon number resolving
detectors. The output state of the control and target qubits is then
final state from these interactions and feed forward
\begin{eqnarray}
 c_0 d_0 |HH \rangle_{ct}+c_0 d_1 |HV \rangle_{ct}+c_1 d_0 |VV
 \rangle_{ct}+c_1 d_1 |VH \rangle_{ct}, 
\end{eqnarray} 
where an additional bit flip was applied to the target qubit if the
ancilla photons state was $|V \rangle$. This final state is the state
that one will expect after a CNOT gate is applied to the initial
control and target qubits. This really shows that our QND-based
parity gates have performed a near deterministic CNOT operation
utilizing only one ancilla qubit (which is not destroyed at the end of
the gate). This represents a huge saving in the physical resources to
implement single photon quantum logic gates. 

\section{Concluding Discussions}

We have shown how  it is possible to create near deterministic two
qubit gates (parity, bell and CNOT) without a huge overhead in ancilla
resources. In fact, an ancilla photon is required only for the CNOT
gate. The key addition to the general linear optical resources  are
weak cross-Kerr nonlinearities and efficient homodyne measurements.
Homodyne measurements are a well established technique frequently used
in  the continuous variable quantum information processing
community. However  weak cross-Kerr nonlinearities are not commonly
used elements within  optical quantum computational devices and as
such it a discussion of the  source and strength of such elements is
required. We will start with a  discussion of the strength of the
nonlinearity as this constraints the  possible physical realisations,
however before this we really need to define  what we mean by {\it
weak} or weak compared with what. Basically it is well  known that
deterministic two qubit gates can be performed if one has access  to a
cross-Kerr nonlinearity that can induce a $\pi$ phase shift directly
between single photon. This leads to a natural definition of weak
nonlinearities,  that is, the use of nonlinear cross-Kerr materials
(when all are taken into  account)  that can not directly induce a
phase shift within an order  of magnitude or several orders of
magnitude of $\pi$. This seems to give an  acceptable functional
definition.

For the parity based gates discussed previously we have established
that the nonlinearity $\theta$ must satisfy the constraint $\alpha_c
\theta^2 \sim 8$ where just to re-emphasise $\alpha_c$ is the
amplitude of the probe beam. Thus due to the weak nature of the
nonlinearity $\theta \ll 1$ we must choose $\alpha_c \sim 10 /
\theta^2$, so for instance if $\theta \sim 10^{-2}$ then $\alpha_c
\geq 10^5$ (which corresponds to a probe beam with mean photon number
$10^{10}$). For a smaller $\theta$ we need a much larger $\alpha_c$.
This puts a natural constraints on $\theta$, since $\alpha_c$ can not
be made arbitrarily large in practice. 

This leads to the question of a mechanism to achieve the weak
cross-Kerr nonlinearity. Natural $\chi^3$ materials have small
nonlinearities  on the order of $10^{-18}$ \cite{kok02}  which would
require lasers with  $\alpha_c \sim 10^{37}$ which is physically
unrealistic. However, systems such as optical fibers \cite{li04},
silica whispering-gallery microresonators \cite{kipp04} and cavity QED
systems \cite{grang98,kimble95}, and EIT \cite{Imamoglu96} are capable
of producing much much larger nonlinearities. For instance
calculations for  EIT systems in NV diamond \cite{munro03} have shown
potential phase shifts of order of magnitude of $\theta=0.01$. With 
$\theta=0.01$ the probe beam must have an amplitude of at least $10^5$
which is physically reasonable with current technology. 

Finally, by using these weak cross-Kerr nonlinearities to aid in the
construction of near deterministic two qubit gates we can build
quantum circuits with far fewer resources than is  known for the
current corresponding linear optical only approaches.It is
straightforward to show in principle that an $n$ qubit computation
requires only of order $n$ single photons sources. This has enormous
implications for the development of single photon quantum computing
and information processing devices and truly indicates the power of a
little nonlinearity.

\vskip 0.5 truecm
\noindent {\em Acknowledgments}: This work was supported by the
European Project RAMBOQ. KN acknowledges support in part from
MPHPT and JSPS. WJM acknowledges support in the form of a JSPS fellowship.


\section{References}
\begin{thebibliography}{99}
\bibitem{KLM} E. Knill, R. Laflamme and G. Milburn, Nature {\bf 409}, 46 (2001).
%
\bibitem{Pittman03} .T. B. Pittman, M.J. Fitch, B.C Jacobs and J.D. Franson, Phys. Rev. A {\bf 68}, 032316 (2003).
%
\bibitem{OBrien03}  J L O'Brien, G J Pryde, A G White, T C Ralph, D Branning, Nature {\bf 426}, 264 (2003).
%
\bibitem{Gasparoni04} S. Gasparoni, J. Pan, P. Walther, T. Rudolph and A. Zeilinger, Phys. Rev. Lett. {\bf 93}, 020504 (2004).
%
\bibitem{scheel03}  Stefan Scheel, Kae Nemoto, W. J. Munro and P. L. Knight, Phys. Rev. A {\bf 68}, 032310 (2003).
%
\bibitem{scheel04} Stefan Scheel, quant-ph/0410014.
%
\bibitem{Eisert04} J. Eisert,  quant-ph/0409156.
%
\bibitem{Vitali00} David Vitali, Mauro Fortunato, and Paolo Tombesiu,  Phys. Rev. Lett. 85, 445-448 (2000)
%
\bibitem{Ottaviani03} Ottaviani et al., Phys. Rev. Lett. 90, 197902 (2003)
%
\bibitem{Imamoglu96}  H.\ Schmidt and A.\ Imamoglu, Optics Letters {\bf 21}, 1936 (1996).
%
\bibitem{Paternotro02}     M. Paternostro, M. S. Kim and B. S. Ham, Physical Review A {\bf 67}, 023811 (2002))
%
\bibitem{munro03}  W. J. Munro, Kae Nemoto, R. G. Beausoleil and T. P. Spiller, in press PRA; quant-ph/0310066.
%
\bibitem{grang98}  P. Grangier, J. A. Levenson and J.-P. Poizat, Nature {\bf 396}, 537 (1998).
%
\bibitem{nemoto04}  Kae Nemoto and W. J. Munro, Phys. Rev. Lett {\bf 93}, 250502 (2004);
%
\bibitem{barrett04}  
S. D. Barrett, P. Kok, Kae Nemoto, R. G. Beausoleil, W. J. Munro and T. P. Spiller, quant-ph/0408117.
%
\bibitem{paris00} M. G. A.  Paris, M. Plenio, D. Jonathan, S. Bose, G.M.  D'Ariano, Phys Lett A {\bf 273}, 153 (2000).
%
\bibitem{DAriano00} G. M. D'Ariano, L. Maccone, M. G. A. Paris and M. F. Sacchi, Phys. Rev. A {\bf 61}. 053817 (2000).
%
\bibitem{Milburn 1984}   G. J. Milburn and D. F. Walls, Phys. Rev. A {\bf 30}, 56 (1984).
%
\bibitem{imot85}  N.\ Imoto, H.\ A.\ Haus, and Y.\ Yamamoto, Phys.\ Rev.\ A {\bf 32}, 2287 (1985).
%
\bibitem{kok02} P. Kok, H. Lee and J. P. Dowling, Phys. Rev. A {\bf 66}, 063814 (2002).
%
\bibitem{li04} X.\ Li, P.\ L. Voss, J.\ E.\ Sharping, and P. Kumar, quant-ph/0402191.
%
\bibitem{kipp04} T.\ J.\ Kippenberg, S.\ M.\ Spillane, and K.\ J.\ Vahala, Phys.\ Rev.\ Lett.\ \textbf{93}, 083904 (2004)..
%
\bibitem{kimble95} 
Q. A. Turchette, C. J. Hood, W. Lange, H. Mabuchi and H. J. Kimble, Phys. Rev. Lett. {\bf 75}, 4710 (1995).
%
\end{thebibliography}
\end{document}